\documentclass[prd,aps,showpacs,nofootinbib,preprint]{revtex4}
\usepackage{graphicx,color,amsmath,amsxtra}
\usepackage{amssymb}
\usepackage[english]{babel}
\usepackage{amsfonts}

\begin{document}

\title{On the necessity of the revisions for the cosmological matter perturbations from the general relativity}

\author{Jiro Matsumoto\footnote{E-mail address: matumoto@th.phys.nagoya-u.ac.jp} }
\affiliation{ Department of Physics, Nagoya University, Nagoya 464-8602, Japan\\ }

\begin{abstract}

The differential equations, which are used for matter perturbations, are usually derived from the Newton gravity 
and the Euler equation in the expanded universe. 
In this paper, by the explicit calculations of metric perturbation theory 
in $\Lambda$CDM model, we derive the general relativistic corrections exactly for the equations 
and show that the differential equations derived from the Newton gravity and the Euler equation
 are valid in the small scale region from non-relativistic matter dominant era onwards. 
However the corrections are not negligible in radiation dominant era. 

\end{abstract}


\maketitle

\section{Introduction \label{SecI}}

In order to explain the accelerating expansion of the present universe, various models of the dark energy 
have been proposed \cite{Amendola et al., Coupled, DGP, f(R)}. 
Cosmological perturbation theory could be a good tool for characterizing each of 
the dark energy models \cite{L. Guzzo et al., T. Narikawa et al.}. 
Because the perturbation theory gives the informations about the difference from the background evolution of the Universe,
which could depend on the characters of the models, we have a possibility to distinguish the characters 
of each model of the dark energy even if they give the identical evolution of the expansion in the Universe. 

The cosmological perturbation theory is often used under the sub-horizon approximation, 
which consists of the two approximations for the gravitational potential in the small scale $a/k \ll 1/H$ 
and in the Hubble scale evolution $1/dt \sim H$, so that the perturbation should be consistent with the Newton gravity. 
The sub-horizon approximation is, however, merely an approximation so that we need to evaluate the deviations 
from the exact solutions in order to justify the approximation. 
While the exact calculations should be executed by using the general relativity, the calculation under the sub-horizon 
approximation gives the results identical with those in the Newton gravity. 
Recently P.~Zhang has investigated the general relativistic corrections from the Newton gravity and has shown 
the corrections could not be neglected in the large scale region \cite{Zhang}. 
As we show in Section \ref{SecIII}, the corrections are negligible in the small scale region 
by the exact evaluation of the general relativistic corrections from matter dominant era onwards. 
But we also show that the corrections are not negligible in radiation dominant era even if we consider 
the matter perturbations in the small scale region. 

We use units of $k_\mathrm{B} = c = \hbar = 1$ and denote the gravitational constant $8 \pi G$ by
${\kappa}^2$ in the following. 

%%%%%%%%%%%%%%%%%%%%%%%%%%%%%%%%%%%%%%%%%%%%%%%%%%%%%%%%%%%%

\section{Cosmological perturbations \label{SecII}}

In this section, we consider the perturbed equations in $\Lambda$CDM model by using the Newtonian gauge.
We assume, however, that the matter components are given by a single barotropic fluid. 
Therefore the following arguments can be applied only in radiation dominant era and from non-relativistic 
matter dominant era onwards.

We begin with the Einstein equation including the cosmological constant:  
\begin{align}
\label{20}
R_{\mu \nu} - \frac{1}{2}g_{\mu \nu} R= - g_{\mu \nu} \Lambda + \kappa ^2 T_{\mu \nu}.
\end{align} 
If we assume that the spacial curvature should vanish, the Friedmann-Lemaitre-Robertson-Walker (FLRW) 
equations are given by substituting the FLRW metric 
$ds^2 = - dt^2 + \delta _{ij} a(t)^2 dx^i dx^j$ into Eq. (\ref{20}): 
\begin{align}
3H^2 = \kappa ^2 \rho + \Lambda, \nonumber \\
-a^2 \delta_{ij} (2 \dot H + 3H^2) = \kappa ^2 a^2 \delta_{ij} p - a^2 \delta _{ij} \Lambda, 
\label{30}
\end{align}
where we assume the energy momentum tensor of perfect fluid, 
$T_{\mu \nu} = pg_{\mu \nu} + (\rho +p)u_{\mu}u_{\nu}, \: u_i =0,\: u_0=-1$, where $u_\mu$ is the four vector satisfying 
$g^{\mu\nu} u_\mu u_\nu = -1$, $\rho$ is the energy density, and $p$ is the pressure. 
By substituting the FLRW metric with perturbation in the Newtonian gauge, 
\begin{align}
\label{60}
ds^2 = (-1+2 \Phi) dt^2 + \delta _{ij} a(t)^2 (1+2 \Psi) dx^i dx^j , 
\end{align}
into the perturbed Einstein equation, 
\begin{align}
\delta R^{\nu}_{\ \mu} - \delta^{\nu}_{\ \mu} \frac{1}{2} \delta R = \kappa ^2 \delta T^{\nu}_{\ \mu}, 
\label{40}
\end{align}
we obtain the following equations in the leading order of the perturbation, 
\begin{align}
 -6H^2 \Phi -2 \frac{k^2}{a^2}\Psi -6H \partial _0 \Psi =&  - \kappa ^2 \delta \rho, 
\label{70} \\
2 \partial _i (H \Phi + \partial _0 \Psi) =& \kappa ^2 (\rho + p) \delta u_i ,
\label{80} \\
a^{-2} \partial _i \partial _j (\Phi - \Psi) =& 0, \quad (i \neq j), 
\label{90} \\
\left( \frac{k^2}{a^2} + \frac{\partial _i \partial _i}{a^2} -2H \partial _0 -4 \dot H -6H^2 \right) \Phi
&- \left( \frac{k^2}{a^2} + \frac{\partial _i \partial _i}{a^2}  + 2 \partial _0 \partial _0 
+ 6H \partial _0 \right) \Psi  \nonumber \\
=& \kappa ^2  \delta p , \quad \left(\mbox{not summed with respect to $i$}\right), 
\label{100} 
\end{align}
where Eqs. (\ref{70}), (\ref{80}), (\ref{90}), and (\ref{100}) express the $(00)$, $(0i)$, 
$(ij)$ for $i \neq j$, and $(ij)$ for $i = j$ components of the Einstein equations, respectively. 
In (\ref{70}) and (\ref{100}), $k$ represents the wave number which appears from the derivative with respect 
to the spacial coordinates ($k^2 = - \partial _j \partial _j$) by the Fourier transformation. 
We treat the energy momentum tensor as that of the perfect fluid so that the perturbations 
of the energy momentum tensor are given by
\begin{align}
\delta T^0_{\ 0} = -\delta \rho, \label{110}\\
\delta T^0_{\ i} = (\rho + p) \delta u_i, \label{120}\\
\delta T^i_{\ 0} = - a^{-2} (\rho + p) \delta u_i, \label{130}\\
\delta T^i_{\ j} = \delta_{ij} \delta p \label{140}, 
\end{align}
where we use $\delta u_0 = \delta g _{00} /2$, which is obtained from the condition $g^{\mu\nu} u_\mu u_\nu = -1$. 
When we consider the scalar perturbation, we decompose $\delta u_i$ as $\partial _i \delta u + \delta u^V_i$, 
where $\partial^i \delta u^V_i =0$ and put $\delta u^V_i =0$ and keep only the scalar part $\delta u$.
Then $\delta u_i$ is expressed as $\delta u_i \equiv \partial _i \delta u$. 
The equation of the perturbation for the matter is given by considering the fluctuation in 
the equation of continuity $\nabla _{\mu} T^{\mu}_{\ \nu} =0$, 
\begin{align}
\delta \dot \rho + 3H(\delta \rho + \delta p) +a^{-2} \partial _i \{ (\rho +p) \delta u_i  \} 
+3 \dot \Psi (\rho + p) =0, \label{150} \\
a^{-3} \partial _0 \{ a^3 (\rho + p) \delta u_i \} + \partial _i \delta p - (\rho + p) \partial _i \Phi =0. \label{160}
\end{align}
Basically, we obtain the time evolution of matter density perturbation $\delta \equiv \delta \rho / \rho$ 
from the above these equations. 
It should be noted, however, there are unknown variables $\delta u_i$ and $\Psi$ and 
therefore we must need to use the other equations such as (\ref{70}), (\ref{80}), and so on.

%%%%%%%%%%%%%%%%%%%%%%%%%%%%%%%%%%%%%%%%%%%%%%%%%%%%%%%%%%%%%%%%%%%%%%%%%%%%%%%%%%%%

\section{Sub-horizon approximation and the exact equation  \label{SecIII}}

\subsection{Deduction of the differential equation}

In the following, we choose $\Psi = \Phi$ in Eq.~(\ref{90}).
If we use the sub-horizon approximation, which is given by $H^2 \Psi$, 
$H \partial_0 \Psi$, $\partial_0^2 \Psi \ll k^2 \Psi / a^2$, Eq.~(\ref{70}) gives, 
\begin{align}
2 \frac{k^2}{a^2}\Psi \simeq  \kappa ^2 \delta \rho.
\label{sub}
\end{align}
This reproduces the Newton potential if we replace $\Psi$, $k^2 / a^2$ and $\delta \rho$ 
by $- \phi$, $- \nabla ^2$ and $\rho$, respectively.
Such an approximation is correct when we consider the case of small sound speed and the region 
that the expansion speed of the Universe is negligible. 
But we should consider the calculation error without any assumptions. 
Thus we now evaluate the correction by the general relativity to the differential equations of 
matter perturbation. 
When we use the sub-horizon approximation and when $w = p = \delta p = 0$, 
we find the differential equation of matter perturbation is given by, 
\begin{equation}
\ddot \delta + 2H \dot \delta - \frac{3}{2} \Omega_m H^2 \delta \simeq 0, 
\label{newton}
\end{equation}
which is given by using Eqs.~(\ref{90}), (\ref{150}), (\ref{160}), and (\ref{sub}). 
Here we define $\dot \delta \equiv \partial _0 \delta$, $\delta \equiv \delta \rho / \rho$, 
and $\Omega _m \equiv \kappa ^2 \rho / 3H^2$. 
The equation (\ref{newton}) is not changed from the equation given by using the Newton gravity 
and the Euler equation \cite{Peebles}. 
If we represent Eq. (\ref{newton}) by using dimensionless variable $N \equiv \ln a$, we obtain
\begin{equation}
\label{subsub}
\frac{d^2 \delta}{dN^2} + \left( \frac{1}{2} - \frac{3}{2}w_\mathrm{eff} \right ) \frac{d \delta}{dN}
 - \frac{3}{2} \Omega _m \delta \simeq 0, 
\end{equation}
where $w_\mathrm{eff} \equiv -1 -2 \dot H / (3H^2)$ is defined by Eqs. in (\ref{30}). 
On the other hand, we use Eqs. (\ref{70}), (\ref{80}), (\ref{90}), and (\ref{150}) 
to obtain the exact equation of matter density perturbation. 
First, Eqs. (\ref{70}) and (\ref{90}) give the one dimensional differential equation for $\dot \Psi$, 
\begin{align}
\dot \Psi =  \frac{ \kappa ^2}{6H} \delta \rho - \bigg ( H + \frac{k^2}{3aH} \bigg ) \Psi. 
\label{190}
\end{align}
Furthermore, this equation (\ref{190}) and Eqs.~(\ref{150}), (\ref{80}), and (\ref{90}) give, 
\begin{align}
\delta \dot \rho =& \bigg \{ -3H(1+c_s ^2) - (1+w) \frac{\kappa ^2 \rho}{2H}
+ \frac{k^2}{3a^2 H} \bigg \} \delta \rho \nonumber \\
& + \bigg \{ 3(1+w)H \rho + (1+w) \frac{\rho k^2}{a^2 H} - \frac{2k^4}{3 a^4 H \kappa ^2} \bigg \} \Psi.
\label{200}
\end{align}
Here the sound velocity $c_s^2$ is defined by $\delta p = c_s ^2 \delta \rho$.
By differentiating Eq. (\ref{200}) with respect to $N$ and eliminating the terms proportional to $\dot \Psi$ and $\Psi$ 
by using (\ref{190}) and (\ref{200}), we find the differential equation expressed only by the variable $\delta \rho$. 
If we use the dimensionless variables $\delta$ and $N=\ln a$, the differential equation is given by 
\begin{align}
\frac{d^2 \delta}{dN^2} + \Bigg \{ 1&+ \frac{3}{2} (1+w) \Omega _m + 3(c_s^2 - w) \nonumber \\
& - \frac{d}{dN} \ln \bigg \vert - \frac{2k^4}{3a^4 H^2 \kappa ^2 \rho} 
+ 3(1+w) \bigg ( 1 + \frac{k^2}{3a^2 H^2} \bigg ) \bigg \vert \Bigg \}
\frac{d \delta}{dN} \nonumber \\
-& \Bigg \{ \frac{k^2}{3 a^2 H^2} (2+ 3w -3 c_s ^2 + 3w_\mathrm{eff}) 
+ 3(w-c_s ^2)- \frac{9}{2}(1+w)(w_\mathrm{eff}-w) \Omega_m \nonumber \\
 &- \bigg ( \frac{k^2}{3a^2 H^2} + 3(w - c_s ^2) - \frac{3}{2}(1+w) \Omega _m \bigg ) 
\frac{d}{dN} \ln \bigg \vert - \frac{2k^4}{3a^4 H^2 \kappa ^2 \rho} \nonumber \\
& \qquad \qquad \qquad \qquad \qquad \qquad \qquad \qquad 
+ 3(1+w) \bigg ( 1 + \frac{k^2}{3a^2 H^2} \bigg ) \bigg \vert \Bigg \}\delta = 0.
\label{210}
\end{align}
This is the exact differential equation of matter perturbation in the Newtonian gauge. 
If we use only small scale approximation, $H^2 \ll k^2 /a^2$ in order to compare Eq.~(\ref{210}) 
with Eq.~(\ref{subsub}), Eq.~(\ref{210}) has the following form:
\begin{align}
\frac{d^2 \delta}{dN^2} -& \left \{ 1+ 3w_\mathrm{eff} +3w -\frac{3}{2}(1+w) \Omega_m 
+ O \left ( \Big ( \frac{k^2}{a^2 H^2} \Big ) ^{-1} \right ) \right \} \frac{d \delta}{dN} \nonumber \\
+& \left \{ \frac{wk^2}{a^2 H^2} - \frac{3}{2}(1+w)(1+3w) \Omega_m 
+ O \left ( \Big ( \frac{k^2}{a^2 H^2} \Big ) ^{-1} \right ) \right \} \delta = 0. 
\label{220}
\end{align}
Here we set $c_s ^2 = w$ for simplicity. 
To derive the above equation (\ref{220}), we used the following expansion, 
\begin{align}
\frac{d}{dN} \ln \bigg \vert - \frac{2k^4}{3a^4 H^2 \kappa ^2 \rho} 
+ 3(1+w) \bigg ( 1 + \frac{k^2}{3a^2 H^2} \bigg ) \bigg \vert & \nonumber \\
=2+ 3w_\mathrm{eff} + 3w +(1+w)(1+3w) \frac{3a^2 \kappa ^2 \rho}{2k^2}
 +& O \left ( \Big ( \frac{k^2}{a^2 H^2} \Big ) ^{-2} \right ).
\label{215}
\end{align}
We need to take $w=0$ to compare with Eq.~(\ref{subsub}) because Eq.~(\ref{subsub}) can be only applied for 
the non-relativistic matter dominant era onwards. 
\begin{align}
\frac{d^2 \delta}{dN^2} + \left \{ \frac{1}{2} - \frac{3}{2}w_\mathrm{eff}
+ O \left ( \Big ( \frac{k^2}{a^2 H^2} \Big ) ^{-1} \right ) \right \} \frac{d \delta}{dN}
+ \left \{  - \frac{3}{2} \Omega_m 
+ O \left ( \Big ( \frac{k^2}{a^2 H^2} \Big ) ^{-1} \right ) \right \} \delta = 0, 
\label{225}
\end{align}
where we use the equation $(1+w) \Omega _{m} = 1+ w_\mathrm{eff}$ given by (\ref{30}). 
Equation (\ref{225}) is a same equation with Eq. (\ref{subsub}) up to $O \left ( \{ k^2/(a^2 H^2) \} ^0 \right )$. 
So that it is shown that the Eq. (\ref{subsub}) is valid in the small scale region $H^2 \ll k^2 /a^2$. 

Whereas the scale we are interested in isn't always deep inside the horizon in radiation dominant era. 
Therefore we need to consider the whole corrections in Eq. (\ref{210}).  
Even if we consider the region deep inside the horizon, we can see that the general relativistic corrections 
are not negligible in radiation dominant era from Eq. (\ref{220}). 
If we express Eq. (\ref{220}) by using time derivative, we obtain 
\begin{align}
\ddot \delta +& \left \{ 2-3w+ O \left ( \Big ( \frac{k^2}{a^2 H^2} \Big ) ^{-1} \right ) \right \} H \dot \delta \nonumber \\
&+ \left \{ \frac{w k^2}{a^2 H^2} - \frac{3}{2}(1+w)(1+3w)\Omega _m + O \left ( \Big ( \frac{k^2}{a^2 H^2} \Big ) ^{-1} \right ) \right \}
\delta = 0.
\label{227}
\end{align}
This equation (\ref{227}) is equivalent to the equation in \cite{Bardeen, Liddle and Lyth} which is derived by 
the different ways in general relativity. 
%%%%%%%%%%%%%%%%%%%%%%%%%%%%%%%%%%%%%%%%%%%%%%%%%%%%%%%%%%%%%%%%%%%%%%%%%%%%%%%%%%%%%%%%%%%%%%

\subsection{Consistency check}

We obtain the differential equation of $\Psi$ from Eqs.~(\ref{190}) and (\ref{200}) in a same way 
as we obtain the differential equation of $\delta$ in the last subsection. 
\begin{align}
\frac{d^2 \Psi}{dN^2} + \left \{ 1 + 3(c_s ^2 - w_\mathrm{eff}) + \frac{3}{2}(1+w) \Omega _m \right \} \frac{d \Psi}{dN}
+ \left \{ \frac{c_s ^2 k^2}{a^2 H^2} + 3(c_s ^2 - w_\mathrm{eff}) \right \} \Psi = 0. 
\label{230}
\end{align}
The differential equation for $\Psi$ can be, however, also obtained by another way. 
Eliminating the terms proportional to $\delta$ and $\Phi$ by using Eqs.~(\ref{70}), (\ref{90}), and (\ref{100}), 
we obtain the following equation, 
\begin{align}
\frac{d^2 \Psi}{dN^2} + \left ( \frac{5}{2} + 3c_s ^2 - \frac{3}{2} w_\mathrm{eff} \right ) \frac{d \Psi}{dN}
+ \left \{ \frac{c_s ^2 k^2}{a^2 H^2} + 3(c_s ^2 - w_\mathrm{eff}) \right \} \Psi = 0.
\label{240}
\end{align}
The obtained expression of Eq.~(\ref{240}) might appear to be different from that in (\ref{230}) but 
if we use the equation $(1+w) \Omega _{m} = 1+ w_\mathrm{eff}$, we can find Eq.~(\ref{240}) is surely identical 
with (\ref{230}). 
This might be considered to be a trivial result but in case that there are non-trivial contribution 
from the diagonal parts of the metric perturbation, the obtained results are not identical with each other. 
%%%%%%%%%%%%%%%%%%%%%%%%%%%%%%%%%%%%%%%%%%%%%%%%%%%%%%%%%%%%%%%%%%%%%%%%%%%%%%

\section{Conclusion \label{SecIV}}

In this paper, we derived the exact differential equation of matter perturbation in the Newtonian gauge. 
This equation (\ref{210}) gives the standard equation of matter density perturbation, 
$\ddot \delta + 2H \dot \delta - 3 \Omega_m H^2 \delta /2 \simeq 0$, from non-relativistic dominant era onwards by 
the small scale approximation $H^2 \ll k^2 /a^2$. 
Whereas in the radiation dominant era, Eq. (\ref{210}) gives the same result in \cite{Bardeen, Liddle and Lyth} at 
the deep inside the horizon. 
So that the Eq. (\ref{210}) can be thought as the generalization of those equations. 
In particular Eq. (\ref{210}) is useful in radiation dominant era because the small scale approximation could not 
work well in radiation dominant era. 
\section*{Acknowledgments}

I am grateful to S.~Nojiri and K.~Bamba for helpful discussions and advices. 
I thank S.~Yokoyama for pointing out a mistake. 
The work is supported in part 
by Global COE Program
of Nagoya University (G07) provided by the Ministry of Education, Culture, 
Sports, Science \& Technology.

%%%%%%%%%%%%%%%%%%%%%%%%%%%%%%%%%
%% thebibliography environment
%%%%%%%%%%%%%%%%%%%%%%%%%%%%%%%%%

\end{document}